\RequirePackage{lineno}
\documentclass[twocolumn,showpacs,aps,prc,superscriptaddress]{revtex4-1}

\usepackage{graphicx}
\usepackage{dcolumn}
\usepackage{amsmath}
\usepackage{supertabular}
\usepackage{multirow}
\usepackage{ulem}
\usepackage{natbib}
\usepackage{lineno}
\begin{document}

\title{Collision Energy Dependence of $p_{\rm t}$ Correlations in Au+Au Collisions at RHIC}

\affiliation{Abilene Christian University, Abilene, Texas   79699}
\affiliation{AGH University of Science and Technology, FPACS, Cracow 30-059, Poland}
\affiliation{Alikhanov Institute for Theoretical and Experimental Physics, Moscow 117218, Russia}
\affiliation{Argonne National Laboratory, Argonne, Illinois 60439}
\affiliation{Brookhaven National Laboratory, Upton, New York 11973}
\affiliation{University of California, Berkeley, California 94720}
\affiliation{University of California, Davis, California 95616}
\affiliation{University of California, Los Angeles, California 90095}
\affiliation{University of California, Riverside, California 92521}
\affiliation{Central China Normal University, Wuhan, Hubei 430079 }
\affiliation{University of Illinois at Chicago, Chicago, Illinois 60607}
\affiliation{Creighton University, Omaha, Nebraska 68178}
\affiliation{Czech Technical University in Prague, FNSPE, Prague 115 19, Czech Republic}
\affiliation{Technische Universit\"at Darmstadt, Darmstadt 64289, Germany}
\affiliation{E\"otv\"os Lor\'and University, Budapest, Hungary H-1117}
\affiliation{Frankfurt Institute for Advanced Studies FIAS, Frankfurt 60438, Germany}
\affiliation{Fudan University, Shanghai, 200433 }
\affiliation{University of Heidelberg, Heidelberg 69120, Germany }
\affiliation{University of Houston, Houston, Texas 77204}
\affiliation{Huzhou University, China}
\affiliation{Indiana University, Bloomington, Indiana 47408}
\affiliation{Institute of Physics, Bhubaneswar 751005, India}
\affiliation{University of Jammu, Jammu 180001, India}
\affiliation{Joint Institute for Nuclear Research, Dubna 141 980, Russia}
\affiliation{Kent State University, Kent, Ohio 44242}
\affiliation{University of Kentucky, Lexington, Kentucky 40506-0055}
\affiliation{Lawrence Berkeley National Laboratory, Berkeley, California 94720}
\affiliation{Lehigh University, Bethlehem, Pennsylvania 18015}
\affiliation{Max-Planck-Institut f\"ur Physik, Munich 80805, Germany}
\affiliation{Michigan State University, East Lansing, Michigan 48824}
\affiliation{National Research Nuclear University MEPhI, Moscow 115409, Russia}
\affiliation{National Institute of Science Education and Research, HBNI, Jatni 752050, India}
\affiliation{National Cheng Kung University, Tainan 70101 }
\affiliation{Nuclear Physics Institute of the CAS, Rez 250 68, Czech Republic}
\affiliation{Ohio State University, Columbus, Ohio 43210}
\affiliation{Institute of Nuclear Physics PAN, Cracow 31-342, Poland}
\affiliation{Panjab University, Chandigarh 160014, India}
\affiliation{Pennsylvania State University, University Park, Pennsylvania 16802}
\affiliation{Institute of High Energy Physics, Protvino 142281, Russia}
\affiliation{Purdue University, West Lafayette, Indiana 47907}
\affiliation{Pusan National University, Pusan 46241, Korea}
\affiliation{Rice University, Houston, Texas 77251}
\affiliation{Rutgers University, Piscataway, New Jersey 08854}
\affiliation{Universidade de S\~ao Paulo, S\~ao Paulo, Brazil 05314-970}
\affiliation{University of Science and Technology of China, Hefei, Anhui 230026}
\affiliation{Shandong University, Qingdao, Shandong 266237}
\affiliation{Shanghai Institute of Applied Physics, Chinese Academy of Sciences, Shanghai 201800}
\affiliation{Southern Connecticut State University, New Haven, Connecticut 06515}
\affiliation{State University of New York, Stony Brook, New York 11794}
\affiliation{Temple University, Philadelphia, Pennsylvania 19122}
\affiliation{Texas A\&M University, College Station, Texas 77843}
\affiliation{University of Texas, Austin, Texas 78712}
\affiliation{Tsinghua University, Beijing 100084}
\affiliation{University of Tsukuba, Tsukuba, Ibaraki 305-8571, Japan}
\affiliation{United States Naval Academy, Annapolis, Maryland 21402}
\affiliation{Valparaiso University, Valparaiso, Indiana 46383}
\affiliation{Variable Energy Cyclotron Centre, Kolkata 700064, India}
\affiliation{Warsaw University of Technology, Warsaw 00-661, Poland}
\affiliation{Wayne State University, Detroit, Michigan 48201}
\affiliation{Yale University, New Haven, Connecticut 06520}

\author{J.~Adam}\affiliation{Creighton University, Omaha, Nebraska 68178}
\author{L.~Adamczyk}\affiliation{AGH University of Science and Technology, FPACS, Cracow 30-059, Poland}
\author{J.~R.~Adams}\affiliation{Ohio State University, Columbus, Ohio 43210}
\author{J.~K.~Adkins}\affiliation{University of Kentucky, Lexington, Kentucky 40506-0055}
\author{G.~Agakishiev}\affiliation{Joint Institute for Nuclear Research, Dubna 141 980, Russia}
\author{M.~M.~Aggarwal}\affiliation{Panjab University, Chandigarh 160014, India}
\author{Z.~Ahammed}\affiliation{Variable Energy Cyclotron Centre, Kolkata 700064, India}
\author{I.~Alekseev}\affiliation{Alikhanov Institute for Theoretical and Experimental Physics, Moscow 117218, Russia}\affiliation{National Research Nuclear University MEPhI, Moscow 115409, Russia}
\author{D.~M.~Anderson}\affiliation{Texas A\&M University, College Station, Texas 77843}
\author{R.~Aoyama}\affiliation{University of Tsukuba, Tsukuba, Ibaraki 305-8571, Japan}
% ...

\collaboration{STAR Collaboration}\noaffiliation

\include{authors}

\date{\today}

\begin{abstract}
 We present two-particle $p_{\rm t}$ correlations as a function of event centrality for Au+Au collisions at $\sqrt{s_{\rm NN}}$ = 7.7, 11.5, 14.5, 19.6, 27, 39, 62.4, and 200 GeV at the Relativistic Heavy Ion Collider using the STAR detector.  These results are compared to previous measurements from CERES at the Super Proton Synchrotron and from ALICE at the Large Hadron Collider.  The data are compared with UrQMD model calculations and with a model based on a Boltzmann-Langevin approach incorporating effects from thermalization.  The relative dynamical correlations for Au+Au collisions at $\sqrt{s_{\rm NN}}$ = 200 GeV show a power law dependence on the number of participant nucleons and agree with the results for Pb+Pb collisions at $\sqrt{s_{\rm NN}} = 2.76~ {\rm TeV}$ from ALICE.  As the collision energy is lowered from $\sqrt{s_{\rm NN}}$ = 200 GeV to 7.7 GeV, the centrality dependence of the relative dynamical correlations departs from the power law behavior observed at the higher collision energies.  In central collisions, the relative dynamical correlations increase with collision energy up to $\sqrt{s_{\rm NN}}$ = 200 GeV in contrast to previous measurements that showed little dependence on the collision energy.
\end{abstract}


\maketitle

\vspace{0.5cm}


The study of event-by-event correlations and fluctuations in global quantities can provide insight 
into the properties of the hot and dense matter created in Au+Au collisions at ultrarelativistic collision energies \cite{PhysRevC.63.064904,PhysRevC.64.041901,PhysRevLett.85.2076,PhysRevLett.85.2689,PhysRevC.60.024901,PhysRevD.60.114028,Heiselberg:2000fk,PhysRevLett.85.2072,Liu2003184,PhysRevC.63.064903,PhysRevC.66.044904,0954-3899-25-3-013,PhysRevLett.81.4816,PhysRevD.65.096008,PhysRevLett.92.162301,PhysRevC.72.044902,ptDifferentialSTAR,ptDifferentialALICE,PhysRevLett.112.032302,PhysRevLett.113.092301,NetKaon}. Correlations of transverse momentum, $p_{\rm t}$, have been proposed as a measure of thermalization \cite{PhysRevLett.92.162301,PhysRevC.85.014905,PhysRevC.95.064901} and as a probe for the critical point of quantum chromodynamics (QCD) \cite{PhysRevD.65.096008,PhysRevLett.102.032301}.   A detailed study of the dependence of two-particle $p_{\rm t}$ correlations on collision energy and centrality may elucidate the effects of thermalization.  If the matter produced in ultrarelativistic collisions passes through a possible QCD critical point, the fluctuations are predicted to increase with respect to a baseline of uncorrelated emission.  A possible signature of the critical point could be non-monotonic behavior of two-particle correlations as a function of collision energy in central collisions.

In this paper we present an experimental study of the collision energy dependence of $p_{\rm t}$ correlations using Au+Au collisions at center of mass energies ranging from $\sqrt{s_{\rm NN}}$ = 7.7 GeV to 200 GeV, taken during the RHIC Beam Energy Scan (BES) using the Solenoidal Tracker at RHIC (STAR).  The 7.7-, 11.5-, 39-, and 62.4-GeV data were taken in 2010.  The 19.6-, 27-, and 200-GeV data were taken in 2011.  The 14.5-GeV data were taken in 2014.  The main detectors used were the Time Projection Chamber (TPC) \cite{ACKERMANN2003624} and the Time of Flight detector (TOF) \cite{RecentTOF}, both located in a solenoidal magnetic field of 0.50 T.  Charged tracks from the TPC with 0.2 GeV/$c$ $\leq p_{\rm t} \leq$ 2.0 GeV/$c$ and $\left|\eta\right|$ $<$ 0.5  were used in this analysis, where $\eta$ is the pseudorapidity. Tracks in the TPC were characterized by the distance of closest approach (DCA), which is the smallest distance between the projection of the track and the measured event vertex. To suppress secondary particles from weak decays, all tracks were required to have a DCA less than 1 cm. Each track was required to have at least 15 measured points and a ratio of the number of measured points to the possible number of measured points greater than 0.52.  Each event was required to have at least one track matched to a TOF hit to minimize pileup.   For each collision energy, events were accepted if they originated from within 1 cm of the center of the focused beam in the plane perpendicular to the beam axis and within 30 cm of the center of STAR along the beam line to achieve uniform detector performance.  The statistical errors were determined by dividing the dataset into five subsets and calculating the observables for each subset. The standard deviation of these observables divided by the square root of the number of subsets was used to calculate the error.  We estimated the systematic errors of the observables by studying the effects of varying the DCA cut from 0.8 to 1.2 cm, varying the acceptance in $\eta$ from $\left| \eta  \right|$ = 0.4 to 0.6, and by varying the lower cut for $p_{\rm t}$ from 0.18 to $0.22~{\rm GeV}/c$.  The average relative systematic errors related to the DCA, the $\eta$ cut, and the lower cut of $p_{\rm t}$ are 1.3\%, 2.7\%, and 4.0\% respectively.

All the data shown are from minimum bias triggers, which were defined as a coincidence of the signals from the Zero Degree Calorimeters (ZDC) \cite{ADLER2003433}, the Vertex Position Detectors (VPD) \cite{LLOPE201423}, and/or the Beam-Beam Counters (BBC) \cite{STAR_BBC}. Table \ref{tab:NumberOfEvents} shows the number of events analyzed at each collision energy.  The centrality bins were defined in terms of a reference multiplicity, which was defined as the number of detected charged particles within an acceptance of $0.5$ $<$ $\left|\eta\right|$ $<$ 1.0.  For the 200 GeV data, this quantity was corrected for the luminosity dependence and the position of the event vertex along the beam axis. This centrality was defined so that the particles used to determine the event centrality did not include the particles used to calculate the $p_{\rm t}$ correlations.  The centrality bins used in this analysis were defined in terms of the fraction of total inelastic cross section.  Specifically the bins were 0-5\% (most central collisions), 5-10\%, 10-20\%, 20-30\%, 30-40\%, 40-50\%, 50-60\%, 60-70\%, and 70-80\% (most peripheral).  The average number of participating nucleons, $N_{\rm part}$, was calculated for each centrality bin at each collision energy using a Monte Carlo Glauber model \cite{GlauberMonteCarlo,GlauberMonteCarloSTAR}.   

\begin{table}
\begin{center}
  \begin{tabular}{c c}
    \hline
    \hline
   $\sqrt{s_{\rm NN}}$ (GeV) & Events (M) \\ 
   \hline
   7.7 & 1.43 \\
   11.5 & 2.46 \\
   14.5 & 12.0 \\
   19.6 & 15.4 \\
   27 & 28.7 \\
   39 & 24.8 \\
   62.4 & 14.9 \\
   200 & 22.2 \\
   \hline
   \hline
\end{tabular}
\caption{\label{tab:NumberOfEvents} Summary of the number of events analyzed in this analysis.}
\end{center}
\end{table}

The results are compared with calculations using the UrQMD model \cite{0954-3899-25-9-308,BASS1998255}.  Version 3.3 of UrQMD with default parameters was used for Au+Au collisions at RHIC energies and version 3.4 was used for Pb+Pb collisions at $\sqrt{s_{\rm NN}}$ = 2.76 TeV.  UrQMD is a hadronic transport model that does not incorporate effects from a deconfined system of quarks and gluons.  For comparison to the STAR results, the STAR acceptance and tracking efficiency were applied with a dependence on particle type, $p_{\rm t}$, collision energy, and centrality.  The detector efficiencies were first obtained from simulations and then applied to the UrQMD results.  For comparison to the ALICE results, the ALICE acceptance was applied to the UrQMD calculations but no efficiency effects were considered.

To characterize the $p_{\rm t}$ correlations, we used the two-particle $p_{\rm t}$ correlation defined as the covariance given by:

\begin{linenomath}\begin{equation}
	\left<\Delta p_{{\rm t},i}, \Delta p_{{\rm t},j}\right> = \frac{1}{N_{\rm events}}\sum_{k=1}^{N_{\rm events}}\frac{C_k}{N_k\left(N_k - 1\right)},
	\label{eq:dptdpt}
\end{equation}\end{linenomath}

where

\begin{linenomath}\begin{equation}
	C_k = \sum_{i=1}^{N_k}\sum_{j=1,j\not=i}^{N_k}\left(p_{{\rm t},i} - \left<\left<p_{\rm t}\right>\right>\right)\left(p_{{\rm t},j} - \left<\left<p_{\rm t}\right>\right>\right).
	\label{eq:Ck}
\end{equation}\end{linenomath}

\noindent
$N_{\rm events}$ is the number of events, $N_k$ is the number of tracks in the $k$th event, and $p_{{\rm t},i}$ is the transverse momentum of the $i$th track in the given event. The event-averaged $p_{\rm t}$ is defined as 

\begin{linenomath}\begin{equation}
	\left<\left<p_{\rm t}\right>\right> = \frac{\sum_{k=1}^{N_{\rm events}}\left<p_{\rm t}\right>_k}{N_{\rm events}},
	\label{eq:aveavept}
\end{equation}\end{linenomath}

\noindent
where $\left<p_{\rm t}\right>_k$ is the average $p_{\rm t}$ of the $k$th event defined as

\begin{linenomath}\begin{equation}
	\left<p_{\rm t}\right>_k = \frac{\sum_{i=1}^{N_k}p_{{\rm t},i}}{N_k}.
	\label{eq:avept}
\end{equation}\end{linenomath}

\noindent
The quantities $\left<\left<p_{\rm t}\right>\right>$ and $\left<\Delta p_{{\rm t},i}, \Delta p_{{\rm t},j}\right>$ were calculated as a function of the reference multiplicity and then averaged over the centrality bin to remove any dependence on the size of the centrality bins \cite{0954-3899-40-10-105104}.

To characterize two-particle $p_{\rm t}$ correlations, we present the relative dynamical correlation, $\sqrt{\left<\Delta p_{{\rm t},i}, \Delta p_{{\rm t},j}\right>}/\left<\left<p_{\rm t}\right>\right>$.
The relative dynamical correlation represents the magnitude of the dynamic fluctuations of the average transverse momentum in units of $\left<\left<p_{\rm t}\right>\right>$ and can be compared directly to the observables used by CERES \cite{Adamova:2003pz} and ALICE \cite{EPJC_74_3077_2014}.

\begin{figure}
	\includegraphics[width=0.50\textwidth]{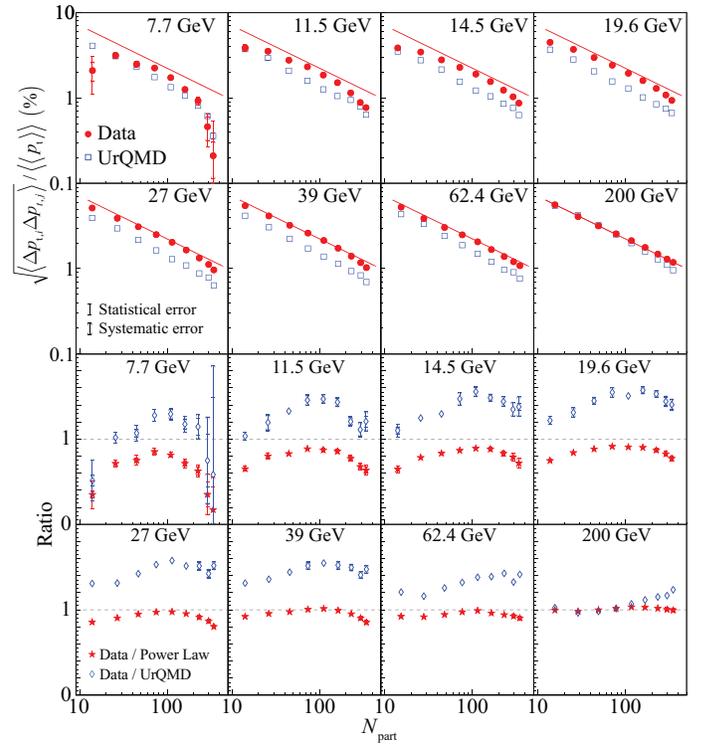}
	\caption{The relative dynamical correlation $\sqrt{\left<\Delta p_{{\rm t},i}, \Delta p_{{\rm t},j}\right>}/\left<\left<p_{\rm t}\right>\right>$ as a function of $N_{\rm part}$ for eight collision energies and UrQMD calculations.  Statistical and systematic errors are shown.  The solid straight lines represent a power law given by $22.32\%/\sqrt{N_{\rm part}}$.  Also shown are the ratios of the measured data to the power law and to UrQMD calculations.}
	\label{fig:sigmapt_general}
\end{figure}

Figure \ref{fig:sigmapt_general} shows the relative dynamical correlation $\sqrt{\left<\Delta p_{{\rm t},i}, \Delta p_{{\rm t},j}\right>}/\left<\left<p_{\rm t}\right>\right>$ as a function of centrality for eight collision energies.  Also shown in this figure are the UrQMD calculations.  The measured relative dynamical correlations for Au+Au collisions at 200 GeV are well reproduced by a power law given by $22.32\%/\sqrt{N_{\rm part}}$.  This power law distribution is also shown for the other seven collision energies.  The relative dynamical correlation distributions deviate from this power law with decreasing collision energy.  Figure \ref{fig:sigmapt_general} also shows the ratio of the measured relative dynamical correlation to the power law distribution observed at 200 GeV and the ratio of the measured values to the UrQMD calculations at each collision energy.

The previous STAR measurements of the relative dynamical correlation at 19.6, 62.4, 130, and 200 GeV \cite{PhysRevC.72.044902} used different acceptance cuts including 0.15 GeV/$c$ $\leq p_{\rm t} \leq$ 2.0 GeV/$c$ and $\left|\eta\right|$ $<$ 1.0 as well as a different centrality definition using detected charged particles with $\left|\eta\right|$ $<$ 0.5.  The previous data at 19.6, 62.4, and 200 GeV are consistent with the current data.

\begin{figure}
	\includegraphics[width=0.50\textwidth]{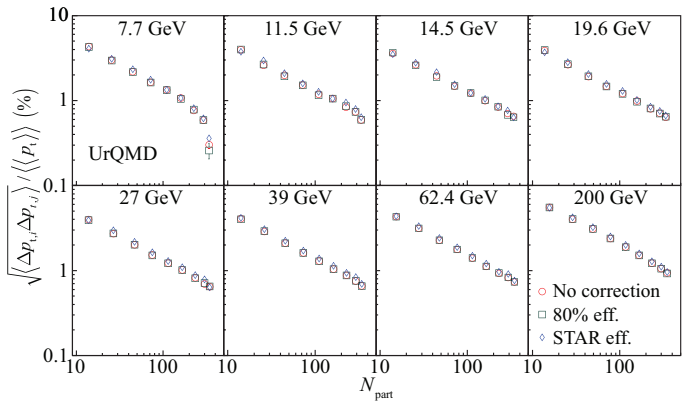}
	\caption{UrQMD calculations for the relative dynamical correlation $\sqrt{\left<\Delta p_{{\rm t},i}, \Delta p_{{\rm t},j}\right>}/\left<\left<p_{\rm t}\right>\right>$ as a function of $N_{\rm part}$ for eight collision energies.  Three cases are shown; UrQMD uncorrected, which has no efficiency corrections, UrQMD 80\% eff., which uses a fixed efficiency of 80\%, and UrQMD, which uses a tracking efficiency that depends on particle type, $p_{\rm t}$, centrality, and collision energy.}
	\label{fig:sigmapt_general_urqmd_compare}
\end{figure}

Figure \ref{fig:sigmapt_general_urqmd_compare} shows the UrQMD results for the relative dynamical correlation for three cases.  The first case is the direct output from the model.  The second case is UrQMD in which the effect of an 80\% constant tracking efficiency was introduced.  The third case is the method used in this paper in which the UrQMD calculations are obtained by introducing the effect of the STAR tracking efficiency, which depends on the particle type, the particle $p_{\rm t}$, the collision energy, and the collision centrality.  These calculations show that the relative dynamical correlation is not sensitive to the efficiency, which allows for the presentation of the experimental results without correction for tracking efficiency.

\begin{figure}
	\includegraphics[width=0.50\textwidth]{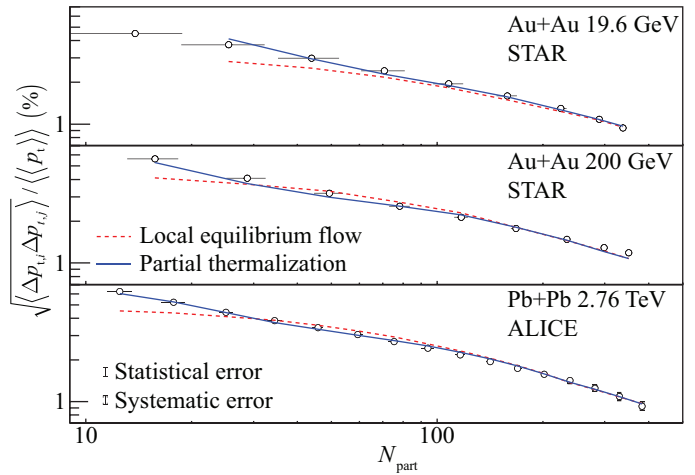}
	\caption{Comparison of a model \cite{PhysRevC.95.064901} incorporating a Boltzmann-Langevin approach to the calculation of thermalization effects for the relative dynamical correlation from Au+Au collisions at $\sqrt{s_{\rm NN}}$ = 19.6 and 200 GeV.  Also shown are model comparisons to results from Pb+Pb collisions at $\sqrt{s_{\rm NN}}$ = 2.76 TeV \cite{EPJC_74_3077_2014}.}
	\label{fig:sigmapt_general_alice_gavin}
\end{figure}

Figure \ref{fig:sigmapt_general_alice_gavin} shows the comparison of a Boltzmann-Langevin approach to the study of equilibration and thermalization effects on two-particle $p_{\rm t}$ correlations \cite{PhysRevC.95.064901}.  The results for local equilibrium flow and partial thermalization are shown compared with current results for 19.6 and 200 GeV collisions as well as the results from Pb+Pb collisions at 2.76 TeV \cite{EPJC_74_3077_2014}.  The local equilibrium flow predictions are realized using a blast wave model including the fluctuation of thermalized flow while a time dependent relaxation time is used to obtain the partial thermalization results.
The authors of Ref. \cite{PhysRevC.95.064901} point out that these comparisons suggest incomplete thermalization in peripheral collisions because they disagree with a local equilibrium flow model.
The agreement in Fig. \ref{fig:sigmapt_general_alice_gavin} of the model calculations for partial thermalization with the measured two-particle $p_{\rm t}$ correlations at all centralities at these three widely-spaced collision energies lends support to this model.

\begin{figure}
	\includegraphics[width=0.50\textwidth]{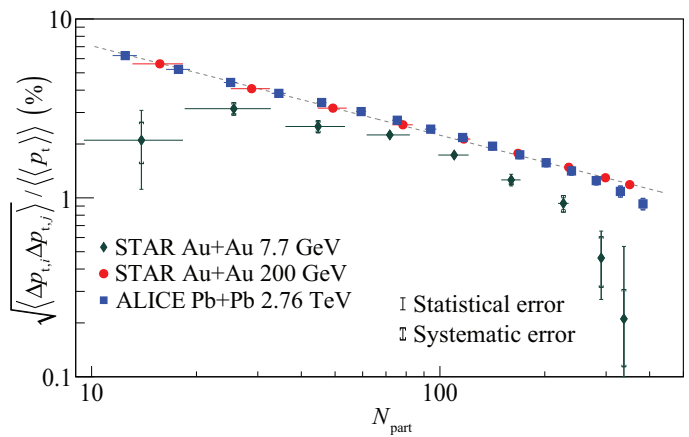}
	\caption{The relative dynamical correlation $\sqrt{\left<\Delta p_{{\rm t},i}, \Delta p_{{\rm t},j}\right>}/\left<\left<p_{\rm t}\right>\right>$ for $\sqrt{s_{\rm NN}}$ = 7.7 GeV and $200~{\rm GeV}$ Au+Au collisions compared with similar results from Pb+Pb collisions at $\sqrt{s_{\rm NN}}$ = 2.76 TeV\cite{EPJC_74_3077_2014}.  The dashed line represents a fit the to data at $\sqrt{s_{\rm NN}}$ = 200 GeV given by $22.32\%/\sqrt{N_{\rm part}}$.  Statistical and systematic errors are shown.}
	\label{fig:sigmapt_general_alice}
\end{figure}

Figure \ref{fig:sigmapt_general_alice} shows the relative dynamical correlation for Au+Au collisions at 7.7 and 200 GeV compared with similar results from Pb+Pb collisions at 2.76 TeV \cite{EPJC_74_3077_2014}.  The ALICE collaboration determined the relative dynamical correlation using tracks with
0.15 GeV/$c$ $\leq p_{\rm t} \leq$ $2.0~{\rm GeV}/c$ and $\left|\eta\right|$ $<$ 0.8.  The results for Au+Au collisions at 200 GeV agree well with the results for Pb+Pb collisions at 2.76 TeV.  The dashed line represents a power law fit to the STAR Au+Au data at 200 GeV of the form $22.32\%/\sqrt {{N_{{\rm{part}}}}}$.  This fit also reproduces the ALICE Pb+Pb results at 2.76 TeV except for the most central collisions.  Not only does the relative dynamical correlation scale as a power law, but it scales as $1/\sqrt{N_{\rm part}}$, adding credence to the idea that the observed particle production comes from uncorrelated sources.  As the collision energy is lowered, the relative dynamical correlation as a function of $N_{\rm part}$ shows a breakdown in this power law scaling as demonstrated by the results for $7.7~{\rm GeV}$ in Fig. \ref{fig:sigmapt_general_alice}.

\begin{figure}
	\includegraphics[width=0.50\textwidth]{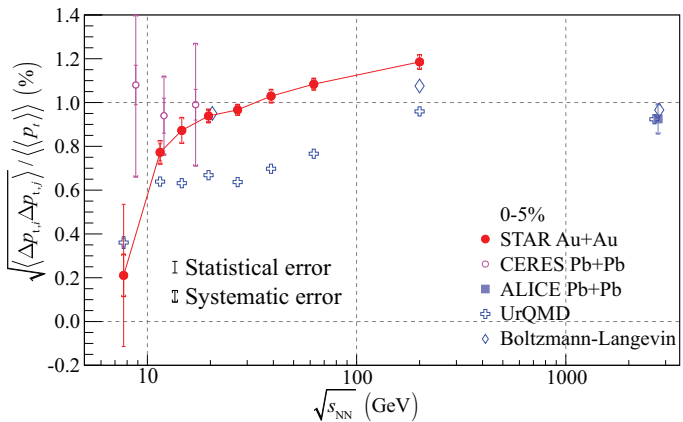}
	\caption{The relative dynamical correlation $\sqrt{\left<\Delta p_{{\rm t},i}, \Delta p_{{\rm t},j}\right>}/\left<\left<p_{\rm t}\right>\right>$ for Au+Au collisions as a function of collision energy for the 0-5\% centrality bin along with results for Pb+Pb from CERES \cite{Adamova:2003pz} and results for Pb+Pb from ALICE \cite{EPJC_74_3077_2014} along with UrQMD calculations and results from Boltzmann-Langevin model calculations \cite{PhysRevC.95.064901}.  The solid line is drawn to guide the eye.  Statistical and systematic errors are shown for the data points.}
	\label{fig:sigmapt_central}
\end{figure}

Figure \ref{fig:sigmapt_central} shows the relative dynamical correlation $\sqrt{\left<\Delta p_{{\rm t},i}, \Delta p_{{\rm t},j}\right>}/\left<\left<p_{\rm t}\right>\right>$ as a function of $\sqrt{s_{\rm NN}}$ for the most central bin (0-5\%). Also shown are the results for Pb+Pb collisions from ALICE \cite{EPJC_74_3077_2014}  and Pb+Pb collisions from CERES \cite{Adamova:2003pz}.  UrQMD calculations are shown as described above.

The data from CERES \cite{Adamova:2003pz} in Fig. \ref{fig:sigmapt_central} are from Pb+Pb collisions at $\sqrt{s_{\rm NN}}$ = 8.7, 12.3 and 17.3 GeV. The CERES results were published using an observable ($\Sigma_{p_{\rm t}}$), which is mathematically identical to $\sqrt{\left<\Delta p_{{\rm t},i}, \Delta p_{{\rm t},j}\right>}/\left<\left<p_{\rm t}\right>\right>$.
STAR had shown previously \cite{PhysRevC.72.044902} that the CERES results taken together with STAR results at 19.6, 62.4, 130, and 200 GeV for 0-5\% centrality indicated that the relative dynamical correlation was constant with collision energy.  The present results for 7.7 GeV to 200 GeV, although in reasonable agreement with the CERES data, show that the relative dynamical correlation decreases at lower collision energy.

Figure \ref{fig:sigmapt_central} also shows the relative dynamical correlation for the 5\% most central collisions from Pb+Pb collisions at 2.76 TeV from ALICE \cite{EPJC_74_3077_2014}.  This result seems to show that the relative dynamical correlation plateaus above 200 GeV.  The relative dynamical correlation at 2.76 TeV is somewhat lower than the value at 200 GeV.  This difference could be partially due to the fact that the 0-5\% centrality bin for Pb+Pb collisions at 2.76 TeV is associated with a somewhat larger value of $N_{\rm part}$ than the value for 200 GeV Au+Au collisions, leading to a lower value of the relative dynamical correlation assuming a $1/ \sqrt{N_{\rm part}}$ scaling.

The UrQMD calculations agree with the measured relative dynamical correlation for Au+Au central collisions at 7.7 GeV and with the relative dynamical correlation for Pb+Pb collisions at 2.76 TeV.  However, the measured relative dynamical correlation increases more than the calculated values from UrQMD as the collision energy is increased from 7.7 GeV to 200 GeV.  Also shown in Fig. \ref{fig:sigmapt_central} are the predictions of the Boltzmann-Langevin calculations \cite{PhysRevC.95.064901} for central collisions of Au+Au at 19.6 and 200 GeV and central collisions of Pb+Pb at 2.76 TeV.  These results show little dependence on the collision energy and agree with the measured results at 19.6 GeV and 2.76 TeV but slightly under-predict the results at 200 GeV.

\begin{figure}
	\includegraphics[width=0.50\textwidth]{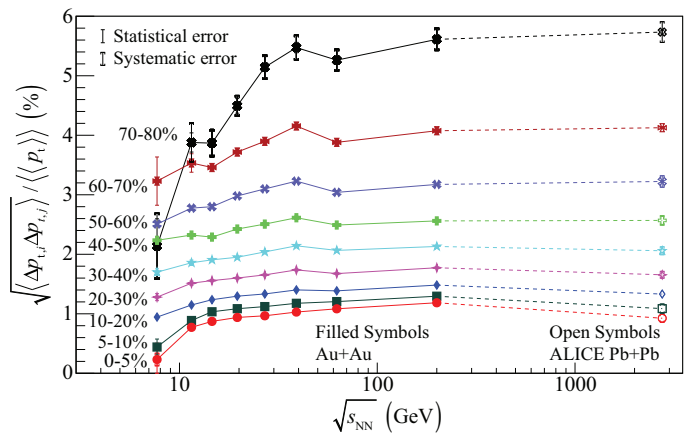}
	\caption{The relative dynamical correlation $\sqrt{\left<\Delta p_{{\rm t},i}, \Delta p_{{\rm t},j}\right>}/\left<\left<p_{\rm t}\right>\right>$ for Au+Au collisions grouped by centrality, compared to Pb+Pb collisions at 2.76 TeV from ALICE \cite{EPJC_74_3077_2014}.  The solid and dashed lines are drawn to guide the eye.  The filled symbols refer to the present Au+Au results while the open symbols refer to Pb+Pb results from ALICE \cite{EPJC_74_3077_2014}.  Statistical and systematic errors are shown.}
	\label{fig:sigmapt_centrality}
\end{figure}

Figure \ref{fig:sigmapt_centrality} shows the relative dynamical correlation for Au+Au collisions along with Pb+Pb collisions from ALICE \cite{EPJC_74_3077_2014} grouped by the collision centrality. The results for the most central collisions (0-5\%) increase as the collision energy is increased and tend to plateau at the highest collision energy.  The mid-central centrality bins show less dependence on the collision energy.  In the most peripheral bin, the relative dynamical correlations again increase with collision energy.  For all centralities, the relative dynamical correlations seem to plateau at the highest collision energy.

In conclusion, we observe a power law scaling of the form $1/\sqrt{N_{\rm part}}$ for the relative dynamical correlation in Au+Au collisions at 200 GeV.  A similar power law scaling had been previously observed in Pb+Pb collisions at 2.76 TeV \cite{EPJC_74_3077_2014} except in the most central collisions.  As the collision energy for Au+Au collisions is decreased to $7.7~{\rm GeV}$, the power law scaling observed at 200 GeV breaks down.  For the most central Au+Au collisions, the relative dynamical correlations increase with collision energy up to 200 GeV and seem to reach a plateau from $200~{\rm GeV}$ to 2.76 TeV.  We observe no non-monotonic scaling of the measured relative dynamical correlation in central collisions.  We observe that two-particle $p_{\rm t}$ correlations show evidence of incomplete thermalization when compared with the Boltzmann-Langevin model in Ref. \cite{PhysRevC.95.064901} in the most peripheral collisions.  New calculations from this model at collision energies below 19.6 GeV would be of interest to better determine the extent of thermalization.  
\section*{Acknowledgments}

We thank the RHIC Operations Group and RCF at BNL, the NERSC Center at LBNL, and the Open Science Grid consortium for providing resources and support.  This work was supported in part by the Office of Nuclear Physics within the U.S. DOE Office of Science, the U.S. National Science Foundation, the Ministry of Education and Science of the Russian Federation, National Natural Science Foundation of China, Chinese Academy of Science, the Ministry of Science and Technology of China and the Chinese Ministry of Education, the National Research Foundation of Korea, Czech Science Foundation and Ministry of Education, Youth and Sports of the Czech Republic, Department of Atomic Energy and Department of Science and Technology of the Government of India, the National Science Centre of Poland, the Ministry  of Science, Education and Sports of the Republic of Croatia, RosAtom of Russia and German Bundesministerium fur Bildung, Wissenschaft, Forschung and Technologie (BMBF) and the Helmholtz Association.

\bibliographystyle{mine}
\bibliography{Correlations_Research}

\begin{thebibliography}{36}%
\makeatletter
\providecommand \@ifxundefined [1]{%
 \@ifx{#1\undefined}
}%
\providecommand \@ifnum [1]{%
 \ifnum #1\expandafter \@firstoftwo
 \else \expandafter \@secondoftwo
 \fi
}%
\providecommand \@ifx [1]{%
 \ifx #1\expandafter \@firstoftwo
 \else \expandafter \@secondoftwo
 \fi
}%
\providecommand \natexlab [1]{#1}%
\providecommand \enquote  [1]{``#1''}%
\providecommand \bibnamefont  [1]{#1}%
\providecommand \bibfnamefont [1]{#1}%
\providecommand \citenamefont [1]{#1}%
\providecommand \href@noop [0]{\@secondoftwo}%
\providecommand \href [0]{\begingroup \@sanitize@url \@href}%
\providecommand \@href[1]{\@@startlink{#1}\@@href}%
\providecommand \@@href[1]{\endgroup#1\@@endlink}%
\providecommand \@sanitize@url [0]{\catcode `\\12\catcode `\$12\catcode
  `\&12\catcode `\#12\catcode `\^12\catcode `\_12\catcode `\%12\relax}%
\providecommand \@@startlink[1]{}%
\providecommand \@@endlink[0]{}%
\providecommand \url  [0]{\begingroup\@sanitize@url \@url }%
\providecommand \@url [1]{\endgroup\@href {#1}{\urlprefix }}%
\providecommand \urlprefix  [0]{URL }%
\providecommand \Eprint [0]{\href }%
\providecommand \doibase [0]{http://dx.doi.org/}%
\providecommand \selectlanguage [0]{\@gobble}%
\providecommand \bibinfo  [0]{\@secondoftwo}%
\providecommand \bibfield  [0]{\@secondoftwo}%
\providecommand \translation [1]{[#1]}%
\providecommand \BibitemOpen [0]{}%
\providecommand \bibitemStop [0]{}%
\providecommand \bibitemNoStop [0]{.\EOS\space}%
\providecommand \EOS [0]{\spacefactor3000\relax}%
\providecommand \BibitemShut  [1]{\csname bibitem#1\endcsname}%
\let\auto@bib@innerbib\@empty
\bibitem [{\citenamefont {Heiselberg}\ and\ \citenamefont
  {Jackson}(2001)}]{PhysRevC.63.064904}%
  \BibitemOpen
  \bibfield  {author} {\bibinfo {author} {H.~Heiselberg}\ and\ \bibinfo
  {author} {A.~D. Jackson},\ }\href {\doibase 10.1103/PhysRevC.63.064904}
  {\bibfield  {journal} {\bibinfo  {journal} {Phys. Rev. C}\ }\textbf {\bibinfo
  {volume} {63}},\ \bibinfo {pages} {064904} (\bibinfo {year}
  {2001})}\BibitemShut {NoStop}%
\bibitem [{\citenamefont {Lin}\ and\ \citenamefont
  {Ko}(2001)}]{PhysRevC.64.041901}%
  \BibitemOpen
  \bibfield  {author} {\bibinfo {author} {Z.-w. Lin}\ and\ \bibinfo {author}
  {C.~M. Ko},\ }\href {\doibase 10.1103/PhysRevC.64.041901} {\bibfield
  {journal} {\bibinfo  {journal} {Phys. Rev. C}\ }\textbf {\bibinfo {volume}
  {64}},\ \bibinfo {pages} {041901} (\bibinfo {year} {2001})}\BibitemShut
  {NoStop}%
\bibitem [{\citenamefont {Jeon}\ and\ \citenamefont
  {Koch}(2000)}]{PhysRevLett.85.2076}%
  \BibitemOpen
  \bibfield  {author} {\bibinfo {author} {S.~Jeon}\ and\ \bibinfo {author}
  {V.~Koch},\ }\href {\doibase 10.1103/PhysRevLett.85.2076} {\bibfield
  {journal} {\bibinfo  {journal} {Phys. Rev. Lett.}\ }\textbf {\bibinfo
  {volume} {85}},\ \bibinfo {pages} {2076} (\bibinfo {year}
  {2000})}\BibitemShut {NoStop}%
\bibitem [{\citenamefont {Bass}\ \emph {\textit{et~al.}}(2000)\citenamefont
  {Bass}, \citenamefont {Danielewicz},\ and\ \citenamefont
  {Pratt}}]{PhysRevLett.85.2689}%
  \BibitemOpen
  \bibfield  {author} {\bibinfo {author} {S.~A. Bass}, \bibinfo {author}
  {P.~Danielewicz},\ and\ \bibinfo {author} {S.~Pratt},\ }\href {\doibase
  10.1103/PhysRevLett.85.2689} {\bibfield  {journal} {\bibinfo  {journal}
  {Phys. Rev. Lett.}\ }\textbf {\bibinfo {volume} {85}},\ \bibinfo {pages}
  {2689} (\bibinfo {year} {2000})}\BibitemShut {NoStop}%
\bibitem [{\citenamefont {Voloshin}\ \emph {\textit{et~al.}}(1999)\citenamefont
  {Voloshin}, \citenamefont {Koch},\ and\ \citenamefont
  {Ritter}}]{PhysRevC.60.024901}%
  \BibitemOpen
  \bibfield  {author} {\bibinfo {author} {S.~A. Voloshin}, \bibinfo {author}
  {V.~Koch},\ and\ \bibinfo {author} {H.~G. Ritter},\ }\href {\doibase
  10.1103/PhysRevC.60.024901} {\bibfield  {journal} {\bibinfo  {journal} {Phys.
  Rev. C}\ }\textbf {\bibinfo {volume} {60}},\ \bibinfo {pages} {024901}
  (\bibinfo {year} {1999})}\BibitemShut {NoStop}%
\bibitem [{\citenamefont {Stephanov}\ \emph
  {\textit{et~al.}}(1999)\citenamefont {Stephanov}, \citenamefont {Rajagopal},\
  and\ \citenamefont {Shuryak}}]{PhysRevD.60.114028}%
  \BibitemOpen
  \bibfield  {author} {\bibinfo {author} {M.~Stephanov}, \bibinfo {author}
  {K.~Rajagopal},\ and\ \bibinfo {author} {E.~Shuryak},\ }\href {\doibase
  10.1103/PhysRevD.60.114028} {\bibfield  {journal} {\bibinfo  {journal} {Phys.
  Rev. D}\ }\textbf {\bibinfo {volume} {60}},\ \bibinfo {pages} {114028}
  (\bibinfo {year} {1999})}\BibitemShut {NoStop}%
\bibitem [{\citenamefont {Heiselberg}(2001)}]{Heiselberg:2000fk}%
  \BibitemOpen
  \bibfield  {author} {\bibinfo {author} {H.~Heiselberg},\ }\href {\doibase
  10.1016/S0370-1573(00)00140-X} {\bibfield  {journal} {\bibinfo  {journal}
  {Phys. Rept.}\ }\textbf {\bibinfo {volume} {351}},\ \bibinfo {pages} {161}
  (\bibinfo {year} {2001})}\BibitemShut {NoStop}%
\bibitem [{\citenamefont {Asakawa}\ \emph {\textit{et~al.}}(2000)\citenamefont
  {Asakawa}, \citenamefont {Heinz},\ and\ \citenamefont
  {M\"uller}}]{PhysRevLett.85.2072}%
  \BibitemOpen
  \bibfield  {author} {\bibinfo {author} {M.~Asakawa}, \bibinfo {author}
  {U.~Heinz},\ and\ \bibinfo {author} {B.~M\"uller},\ }\href {\doibase
  10.1103/PhysRevLett.85.2072} {\bibfield  {journal} {\bibinfo  {journal}
  {Phys. Rev. Lett.}\ }\textbf {\bibinfo {volume} {85}},\ \bibinfo {pages}
  {2072} (\bibinfo {year} {2000})}\BibitemShut {NoStop}%
\bibitem [{\citenamefont {Liu}\ and\ \citenamefont
  {Trainor}(2003)}]{Liu2003184}%
  \BibitemOpen
  \bibfield  {author} {\bibinfo {author} {Q.~Liu}\ and\ \bibinfo {author}
  {T.~A. Trainor},\ }\href {\doibase 10.1016/j.physletb.2003.06.055} {\bibfield
   {journal} {\bibinfo  {journal} {Phys. Lett. B}\ }\textbf {\bibinfo {volume}
  {567}},\ \bibinfo {pages} {184} (\bibinfo {year} {2003})}\BibitemShut
  {NoStop}%
\bibitem [{\citenamefont {Shuryak}\ and\ \citenamefont
  {Stephanov}(2001)}]{PhysRevC.63.064903}%
  \BibitemOpen
  \bibfield  {author} {\bibinfo {author} {E.~V. Shuryak}\ and\ \bibinfo
  {author} {M.~A. Stephanov},\ }\href {\doibase 10.1103/PhysRevC.63.064903}
  {\bibfield  {journal} {\bibinfo  {journal} {Phys. Rev. C}\ }\textbf {\bibinfo
  {volume} {63}},\ \bibinfo {pages} {064903} (\bibinfo {year}
  {2001})}\BibitemShut {NoStop}%
\bibitem [{\citenamefont {Pruneau}\ \emph {\textit{et~al.}}(2002)\citenamefont
  {Pruneau}, \citenamefont {Gavin},\ and\ \citenamefont
  {Voloshin}}]{PhysRevC.66.044904}%
  \BibitemOpen
  \bibfield  {author} {\bibinfo {author} {C.~Pruneau}, \bibinfo {author}
  {S.~Gavin},\ and\ \bibinfo {author} {S.~Voloshin},\ }\href {\doibase
  10.1103/PhysRevC.66.044904} {\bibfield  {journal} {\bibinfo  {journal} {Phys.
  Rev. C}\ }\textbf {\bibinfo {volume} {66}},\ \bibinfo {pages} {044904}
  (\bibinfo {year} {2002})}\BibitemShut {NoStop}%
\bibitem [{\citenamefont {Bass}\ \emph {\textit{et~al.}}(1999)\citenamefont
  {Bass}, \citenamefont {Gyulassy}, \citenamefont {St{\"o}cker},\ and\
  \citenamefont {Greiner}}]{0954-3899-25-3-013}%
  \BibitemOpen
  \bibfield  {author} {\bibinfo {author} {S.~A. Bass}, \bibinfo {author}
  {M.~Gyulassy}, \bibinfo {author} {H.~St{\"o}cker},\ and\ \bibinfo {author}
  {W.~Greiner},\ }\href {http://stacks.iop.org/0954-3899/25/i=3/a=013}
  {\bibfield  {journal} {\bibinfo  {journal} {J. Phys. G}\ }\textbf {\bibinfo
  {volume} {25}},\ \bibinfo {pages} {R1} (\bibinfo {year} {1999})}\BibitemShut
  {NoStop}%
\bibitem [{\citenamefont {Stephanov}\ \emph
  {\textit{et~al.}}(1998)\citenamefont {Stephanov}, \citenamefont {Rajagopal},\
  and\ \citenamefont {Shuryak}}]{PhysRevLett.81.4816}%
  \BibitemOpen
  \bibfield  {author} {\bibinfo {author} {M.~Stephanov}, \bibinfo {author}
  {K.~Rajagopal},\ and\ \bibinfo {author} {E.~Shuryak},\ }\href {\doibase
  10.1103/PhysRevLett.81.4816} {\bibfield  {journal} {\bibinfo  {journal}
  {Phys. Rev. Lett.}\ }\textbf {\bibinfo {volume} {81}},\ \bibinfo {pages}
  {4816} (\bibinfo {year} {1998})}\BibitemShut {NoStop}%
\bibitem [{\citenamefont {Stephanov}(2002)}]{PhysRevD.65.096008}%
  \BibitemOpen
  \bibfield  {author} {\bibinfo {author} {M.~Stephanov},\ }\href {\doibase
  10.1103/PhysRevD.65.096008} {\bibfield  {journal} {\bibinfo  {journal} {Phys.
  Rev. D}\ }\textbf {\bibinfo {volume} {65}},\ \bibinfo {pages} {096008}
  (\bibinfo {year} {2002})}\BibitemShut {NoStop}%
\bibitem [{\citenamefont {Gavin}(2004)}]{PhysRevLett.92.162301}%
  \BibitemOpen
  \bibfield  {author} {\bibinfo {author} {S.~Gavin},\ }\href {\doibase
  10.1103/PhysRevLett.92.162301} {\bibfield  {journal} {\bibinfo  {journal}
  {Phys. Rev. Lett.}\ }\textbf {\bibinfo {volume} {92}},\ \bibinfo {pages}
  {162301} (\bibinfo {year} {2004})}\BibitemShut {NoStop}%
\bibitem [{\citenamefont {Adams}\ \emph {\textit{et~al.}}(2005)\citenamefont
  {Adams} \emph {\textit{et~al.}}}]{PhysRevC.72.044902}%
  \BibitemOpen
  \bibfield  {author} {\bibinfo {author} {J.~Adams} \textit{et~al.} (\bibinfo
  {collaboration} {STAR Collaboration}),\ }\href {\doibase
  10.1103/PhysRevC.72.044902} {\bibfield  {journal} {\bibinfo  {journal} {Phys.
  Rev. C}\ }\textbf {\bibinfo {volume} {72}},\ \bibinfo {pages} {044902}
  (\bibinfo {year} {2005})}\BibitemShut {NoStop}%
\bibitem [{\citenamefont {Agakishiev}\ \emph
  {\textit{et~al.}}(2011)\citenamefont {Agakishiev} \emph
  {\textit{et~al.}}}]{ptDifferentialSTAR}%
  \BibitemOpen
  \bibfield  {author} {\bibinfo {author} {G.~Agakishiev} \textit{et~al.}
  (\bibinfo {collaboration} {STAR Collaboration}),\ }\href@noop {} {\bibfield
  {journal} {\bibinfo  {journal} {Phys. Lett. B}\ }\textbf {\bibinfo {volume}
  {704}},\ \bibinfo {pages} {467} (\bibinfo {year} {2011})}\BibitemShut
  {NoStop}%
\bibitem [{\citenamefont {Adam}\ \emph {\textit{et~al.}}(2017)\citenamefont
  {Adam} \emph {\textit{et~al.}}}]{ptDifferentialALICE}%
  \BibitemOpen
  \bibfield  {author} {\bibinfo {author} {J.~Adam} \textit{et~al.} (\bibinfo
  {collaboration} {ALICE Collaboration}),\ }\href@noop {} {\bibfield  {journal}
  {\bibinfo  {journal} {Phys. Rev. Lett.}\ }\textbf {\bibinfo {volume} {118}},\
  \bibinfo {pages} {162302} (\bibinfo {year} {2017})}\BibitemShut {NoStop}%
\bibitem [{\citenamefont {Adamczyk}\ \emph
  {\textit{et~al.}}(2014{\natexlab{a}})\citenamefont {Adamczyk} \emph
  {\textit{et~al.}}}]{PhysRevLett.112.032302}%
  \BibitemOpen
  \bibfield  {author} {\bibinfo {author} {L.~Adamczyk} \textit{et~al.}
  (\bibinfo {collaboration} {STAR Collaboration}),\ }\href {\doibase
  10.1103/PhysRevLett.112.032302} {\bibfield  {journal} {\bibinfo  {journal}
  {Phys. Rev. Lett.}\ }\textbf {\bibinfo {volume} {112}},\ \bibinfo {pages}
  {032302} (\bibinfo {year} {2014}{\natexlab{a}})}\BibitemShut {NoStop}%
\bibitem [{\citenamefont {Adamczyk}\ \emph
  {\textit{et~al.}}(2014{\natexlab{b}})\citenamefont {Adamczyk} \emph
  {\textit{et~al.}}}]{PhysRevLett.113.092301}%
  \BibitemOpen
  \bibfield  {author} {\bibinfo {author} {L.~Adamczyk} \textit{et~al.}
  (\bibinfo {collaboration} {STAR Collaboration}),\ }\href {\doibase
  10.1103/PhysRevLett.113.092301} {\bibfield  {journal} {\bibinfo  {journal}
  {Phys. Rev. Lett.}\ }\textbf {\bibinfo {volume} {113}},\ \bibinfo {pages}
  {092301} (\bibinfo {year} {2014}{\natexlab{b}})}\BibitemShut {NoStop}%
\bibitem [{\citenamefont {Adamczyk}\ \emph {\textit{et~al.}}(2018)\citenamefont
  {Adamczyk} \emph {\textit{et~al.}}}]{NetKaon}%
  \BibitemOpen
  \bibfield  {author} {\bibinfo {author} {L.~Adamczyk} \textit{et~al.}
  (\bibinfo {collaboration} {STAR Collaboration}),\ }\href@noop {} {\bibfield
  {journal} {\bibinfo  {journal} {Phys. Lett. B}\ }\textbf {\bibinfo {volume}
  {785}},\ \bibinfo {pages} {551} (\bibinfo {year} {2018})}\BibitemShut
  {NoStop}%
\bibitem [{\citenamefont {Gavin}\ and\ \citenamefont
  {Moschelli}(2012)}]{PhysRevC.85.014905}%
  \BibitemOpen
  \bibfield  {author} {\bibinfo {author} {S.~Gavin}\ and\ \bibinfo {author}
  {G.~Moschelli},\ }\href {\doibase 10.1103/PhysRevC.85.014905} {\bibfield
  {journal} {\bibinfo  {journal} {Phys. Rev. C}\ }\textbf {\bibinfo {volume}
  {85}},\ \bibinfo {pages} {014905} (\bibinfo {year} {2012})}\BibitemShut
  {NoStop}%
\bibitem [{\citenamefont {Gavin}\ \emph {\textit{et~al.}}(2017)\citenamefont
  {Gavin}, \citenamefont {Moschelli},\ and\ \citenamefont
  {Zin}}]{PhysRevC.95.064901}%
  \BibitemOpen
  \bibfield  {author} {\bibinfo {author} {S.~Gavin}, \bibinfo {author}
  {G.~Moschelli},\ and\ \bibinfo {author} {C.~Zin},\ }\href {\doibase
  10.1103/PhysRevC.95.064901} {\bibfield  {journal} {\bibinfo  {journal} {Phys.
  Rev. C}\ }\textbf {\bibinfo {volume} {95}},\ \bibinfo {pages} {064901}
  (\bibinfo {year} {2017})}\BibitemShut {NoStop}%
\bibitem [{\citenamefont {Stephanov}(2009)}]{PhysRevLett.102.032301}%
  \BibitemOpen
  \bibfield  {author} {\bibinfo {author} {M.~A. Stephanov},\ }\href {\doibase
  10.1103/PhysRevLett.102.032301} {\bibfield  {journal} {\bibinfo  {journal}
  {Phys. Rev. Lett.}\ }\textbf {\bibinfo {volume} {102}},\ \bibinfo {pages}
  {032301} (\bibinfo {year} {2009})}\BibitemShut {NoStop}%
\bibitem [{\citenamefont {Ackermann}\ \emph
  {\textit{et~al.}}(2003)\citenamefont {Ackermann} \emph
  {\textit{et~al.}}}]{ACKERMANN2003624}%
  \BibitemOpen
  \bibfield  {author} {\bibinfo {author} {K.~H. Ackermann} \textit{et~al.}
  (\bibinfo {collaboration} {STAR Collaboration}),\ }\href {\doibase
  https://doi.org/10.1016/S0168-9002(02)01960-5} {\bibfield  {journal}
  {\bibinfo  {journal} {Nucl. Instrum. Methods A}\ }\textbf {\bibinfo {volume}
  {499}},\ \bibinfo {pages} {624} (\bibinfo {year} {2003})}\BibitemShut
  {NoStop}%
\bibitem [{\citenamefont {Llope}(2012)}]{RecentTOF}%
  \BibitemOpen
  \bibfield  {author} {\bibinfo {author} {W.~J. Llope},\ }\href@noop {}
  {\bibfield  {journal} {\bibinfo  {journal} {Nucl. Instrum. Methods A}\
  }\textbf {\bibinfo {volume} {661}},\ \bibinfo {pages} {S110} (\bibinfo {year}
  {2012})}\BibitemShut {NoStop}%
\bibitem [{\citenamefont {Adler}\ \emph {\textit{et~al.}}(2003)\citenamefont
  {Adler}, \citenamefont {Denisov}, \citenamefont {Garcia}, \citenamefont
  {Murray}, \citenamefont {Strobele},\ and\ \citenamefont
  {White}}]{ADLER2003433}%
  \BibitemOpen
  \bibfield  {author} {\bibinfo {author} {C.~Adler}, \bibinfo {author}
  {A.~Denisov}, \bibinfo {author} {E.~Garcia}, \bibinfo {author} {M.~Murray},
  \bibinfo {author} {H.~Strobele},\ and\ \bibinfo {author} {S.~White},\ }\href
  {\doibase https://doi.org/10.1016/j.nima.2003.08.112} {\bibfield  {journal}
  {\bibinfo  {journal} {Nucl. Instrum. Methods A}\ }\textbf {\bibinfo {volume}
  {499}},\ \bibinfo {pages} {433} (\bibinfo {year} {2003})}\BibitemShut
  {NoStop}%
\bibitem [{\citenamefont {Llope}\ \emph {\textit{et~al.}}(2014)\citenamefont
  {Llope} \emph {\textit{et~al.}}}]{LLOPE201423}%
  \BibitemOpen
  \bibfield  {author} {\bibinfo {author} {W.~J. Llope} \textit{et~al.},\ }\href
  {\doibase https://doi.org/10.1016/j.nima.2014.04.080} {\bibfield  {journal}
  {\bibinfo  {journal} {Nucl. Instrum. Methods A}\ }\textbf {\bibinfo {volume}
  {759}},\ \bibinfo {pages} {23} (\bibinfo {year} {2014})}\BibitemShut
  {NoStop}%
\bibitem [{\citenamefont {Whitten}(2008)}]{STAR_BBC}%
  \BibitemOpen
  \bibfield  {author} {\bibinfo {author} {C.~A. Whitten} (\bibinfo
  {collaboration} {STAR Collaboration}),\ }\href@noop {} {\bibfield  {journal}
  {\bibinfo  {journal} {AIP Conference Proceedings}\ }\textbf {\bibinfo
  {volume} {980}},\ \bibinfo {pages} {390} (\bibinfo {year}
  {2008})}\BibitemShut {NoStop}%
\bibitem [{\citenamefont {Miller}\ \emph {\textit{et~al.}}(2007)\citenamefont
  {Miller}, \citenamefont {Reygers}, \citenamefont {Sanders},\ and\
  \citenamefont {Steinberg}}]{GlauberMonteCarlo}%
  \BibitemOpen
  \bibfield  {author} {\bibinfo {author} {M.~L. Miller}, \bibinfo {author}
  {K.~Reygers}, \bibinfo {author} {S.~J. Sanders},\ and\ \bibinfo {author}
  {P.~Steinberg},\ }\href@noop {} {\bibfield  {journal} {\bibinfo  {journal}
  {Ann. Rev. Nucl. Part. Sci.}\ }\textbf {\bibinfo {volume} {57}},\ \bibinfo
  {pages} {205} (\bibinfo {year} {2007})}\BibitemShut {NoStop}%
\bibitem [{\citenamefont {Abelev}\ \emph {\textit{et~al.}}(2009)\citenamefont
  {Abelev} \emph {\textit{et~al.}}}]{GlauberMonteCarloSTAR}%
  \BibitemOpen
  \bibfield  {author} {\bibinfo {author} {B.~I. Abelev} \textit{et~al.}
  (\bibinfo {collaboration} {STAR Collaboration}),\ }\href@noop {} {\bibfield
  {journal} {\bibinfo  {journal} {Phys. Rev. C}\ }\textbf {\bibinfo {volume}
  {79}},\ \bibinfo {pages} {034909} (\bibinfo {year} {2009})}\BibitemShut
  {NoStop}%
\bibitem [{\citenamefont {Bleicher}\ \emph {\textit{et~al.}}(1999)\citenamefont
  {Bleicher} \emph {\textit{et~al.}}}]{0954-3899-25-9-308}%
  \BibitemOpen
  \bibfield  {author} {\bibinfo {author} {M.~Bleicher} \textit{et~al.},\ }\href
  {http://stacks.iop.org/0954-3899/25/i=9/a=308} {\bibfield  {journal}
  {\bibinfo  {journal} {J. Phys. G}\ }\textbf {\bibinfo {volume} {25}},\
  \bibinfo {pages} {1859} (\bibinfo {year} {1999})}\BibitemShut {NoStop}%
\bibitem [{\citenamefont {Bass}\ \emph {\textit{et~al.}}(1998)\citenamefont
  {Bass} \emph {\textit{et~al.}}}]{BASS1998255}%
  \BibitemOpen
  \bibfield  {author} {\bibinfo {author} {S.~Bass} \textit{et~al.},\ }\href
  {\doibase https://doi.org/10.1016/S0146-6410(98)00058-1} {\bibfield
  {journal} {\bibinfo  {journal} {Prog. Part. Nucl. Phys.}\ }\textbf {\bibinfo
  {volume} {41}},\ \bibinfo {pages} {255} (\bibinfo {year} {1998})}\BibitemShut
  {NoStop}%
\bibitem [{\citenamefont {Luo}\ \emph {\textit{et~al.}}(2013)\citenamefont
  {Luo}, \citenamefont {Xu}, \citenamefont {Mohanty},\ and\ \citenamefont
  {Xu}}]{0954-3899-40-10-105104}%
  \BibitemOpen
  \bibfield  {author} {\bibinfo {author} {X.~Luo}, \bibinfo {author} {J.~Xu},
  \bibinfo {author} {B.~Mohanty},\ and\ \bibinfo {author} {N.~Xu},\ }\href
  {http://stacks.iop.org/0954-3899/40/i=10/a=105104} {\bibfield  {journal}
  {\bibinfo  {journal} {J. Phys. G}\ }\textbf {\bibinfo {volume} {40}},\
  \bibinfo {pages} {105104} (\bibinfo {year} {2013})}\BibitemShut {NoStop}%
\bibitem [{\citenamefont {Adamova}\ \emph {\textit{et~al.}}(2003)\citenamefont
  {Adamova} \emph {\textit{et~al.}}}]{Adamova:2003pz}%
  \BibitemOpen
  \bibfield  {author} {\bibinfo {author} {D.~Adamova} \textit{et~al.} (\bibinfo
  {collaboration} {CERES Collaboration}),\ }\href@noop {} {\bibfield  {journal}
  {\bibinfo  {journal} {Nucl. Phys. A}\ }\textbf {\bibinfo {volume} {727}},\
  \bibinfo {pages} {97} (\bibinfo {year} {2003})}\BibitemShut {NoStop}%
\bibitem [{\citenamefont {Abelev}\ \emph {\textit{et~al.}}(2014)\citenamefont
  {Abelev} \emph {\textit{et~al.}}}]{EPJC_74_3077_2014}%
  \BibitemOpen
  \bibfield  {author} {\bibinfo {author} {B.~Abelev} \textit{et~al.} (\bibinfo
  {collaboration} {ALICE Collaboration}),\ }\href {\doibase
  10.1140/epjc/s10052-014-3077-y} {\bibfield  {journal} {\bibinfo  {journal}
  {Eur. Phys. J. C}\ }\textbf {\bibinfo {volume} {74}},\ \bibinfo {pages}
  {3077} (\bibinfo {year} {2014})}\BibitemShut {NoStop}%
\end{thebibliography}%

\end{document}